\documentclass[aps,showpacs,floats]{revtex4}

\usepackage{graphicx}

\begin{document}

\title {Measuring High Energy Neutrino-Nucleon Cross Sections With Future Neutrino Telescopes}
\author{Dan Hooper}
\affiliation{
Department of Physics, University of Wisconsin,
1150 University Avenue, 
Madison, WI 53706}

\begin{abstract}
Next generation kilometer-scale neutrino telescopes, such as ICECUBE, can test standard model predictions for neutrino-nucleon cross sections at energies well beyond the reach of collider experiments.  At energies near a PeV and higher, the Earth becomes opaque to neutrinos.  At these energies, the ratio of upgoing and downgoing events can be used to measure the total neutrino-nucleon cross section given the presence of an adequate high energy neutrino flux.
\end{abstract}


\maketitle
\thispagestyle{empty}

The ability to measure neutrino-nucleon cross sections beyond the energies accessible at colliders will be a valuable tool, capable of addressing multiple open questions in particle physics.  First, the high energy cross section provides information on small-x parton distribution functions \cite{neu}.  Second, physics beyond the standard model can be constrained including scenarios with low scale quantum gravity \cite{ED}.

Present and next generation high energy neutrino telescopes consist of strings of photo-multiplier tubes distributed throughout a Cerenkov medium such as water or ice.  Neutrinos are detected from the hadronic or electromagnetic showers generated in the interactions which take place within the detector volume or from charged lepton tracks generated within the lepton's range of the detector in charged current interactions.  The calculations in this paper take into accout only shower events.  For a review of high energy neutrino astronomy see \cite{icecube,amanda}.

The interaction length for a particle traveling through a number density of targets, $n$ is 
\begin{eqnarray}
l=(\sigma n )^{-1}.
\end{eqnarray}
This length is equal to the diameter of the Earth for a cross section of
\begin{eqnarray}
\sigma=(2 R_{\rm{Earth}}n)^{-1}\sim 2 \times 10^{-7}\rm{mb},  
\end{eqnarray}
which is predicted (but not yet measured) to occur near $E_{\nu} \sim 100$ TeV for neutrino-nucleon interactions.  The fraction of neutrinos which are absorbed by the Earth is a function of cross section.  This can be expressed independently of the flux, as the ratio of downgoing events to upgoing events, at a given energy or in a given energy range.  Figure 1 shows this relationship.  The simulation used for this calculation considered a  detector located at a depth of 1.2 to 2.4 km beneath the Earth's surface.  The Earth was taken to have a core of radius 2500 km and density $11,000 \, \rm{kg}/\rm{m}^3$ and a 2 km later of ice/water along the surface.

Figure 1 shows that below $\sim10^{-7}$ mb, the ratio of downgoing to upgoing events changes between 1 and 1.2$\,$fairly slowly and may be difficult to observe.  Conversely, above $\sim 10^{-4}$ mb, the ratio grows rapidly and well above the number of events that we may expect to observe, making a measurement difficult due to poor statistics.  For this reason, this technique is most well suited for energies within this range of cross sections.  Above these energies, ground-level fluorescence cosmic ray detectors, such as EUSO and OWL, may be able to make accurate measurements \cite{kusenko}.

To predict how effectively we will be able to measure high energy neutrino cross sections, knowledge of the flux of neutrinos at the relevant energies is needed.  A variety of such fluxes have been discussed in the literature.  These include, but are not limited to, neutrinos from compact objects such as Gamma-Ray Bursts (GRB) \cite{grb} and Active Galactic Nuclei (AGN) \cite{agn}, cosmogenic neutrinos generated by cosmic rays scattering off of the photon background \cite{cosmogenic} and top-down scenarios where neutrinos are generated in mechanisms such as the decay of supermassive particles, topological defects or primordial black holes \cite{td}.  In my calculations, I considered four cases.  First, the Waxman-Bahcall flux for transparent sources of cosmic rays.  This is a conservative choice because a more opaque source will yield higher neutrino fluxes.  This flux is given by $E_{\nu}^2 dN_{\nu}/dE_{\nu}=10^{-8}\, \rm{GeV}\, \rm{cm}^{-2}\, \rm{s}^{-1}\, \rm{sr}^{-1}$ for each of $\nu_e, \nu_{\mu}$ and $\bar{\nu_{\mu}}$ \cite{wb}.  Secondly, I used the present flux limit for AMANDA-B10 of $E_{\nu}^2 dN_{\nu}/dE_{\nu}=9 \times 10^{-7} \,\rm{GeV}\, \rm{cm}^{-2}\, \rm{s}^{-1}\, \rm{sr}^{-1}$ for each of $\nu_{\mu}$ and $\bar{\nu_{\mu}}$ \cite{amanda}.  Finally, I used a fluxes of $E_{\nu} dN_{\nu}/dE_{\nu}=6.3 \times 10^{-12} \, \rm{cm}^{-2}\, \rm{s}^{-1}\, \rm{sr}^{-1}$ and $E_{\nu} dN_{\nu}/dE_{\nu}=5.7 \times 10^{-10} \, \rm{cm}^{-2}\, \rm{s}^{-1}\, \rm{sr}^{-1}$ for each of $\nu_{\mu}$ and $\bar{\nu_{\mu}}$ for comparison.  The last two fluxes are normalized to the same number of events between 1 PeV and 1 EeV as for the Waxman-Bahcall flux and the AMANDA-B10 limit, respectively.

\begin{figure}[thb]
\centering\leavevmode
\includegraphics[width=2.5in]{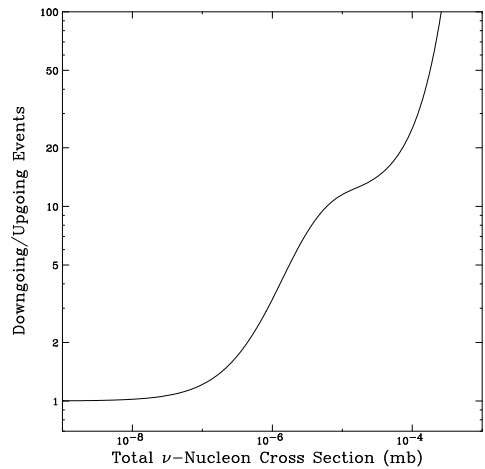}
\caption{The ratio of downgoing to upgoing events in a neutrino telescope as a function of neutrino-nucleon cross section (mb).}
\label{figure}
\end{figure}

The distributions of upgoing and downgoing events are each fit by Poisson statistics.  The ratio of these rates is, therefore, described by a binomial distribution.  Using the astrophysics convention of 84.13\% confidence upper and lower limits containing a 68.27\% confidence interval, error bars can be fit for any pair of values for the number of upgoing and downgoing events \cite{gehrels}.  Figures 2 through 5 show the ability of a cubic kilometer neutrino telescope to constrain the total neutrino-nucleon cross section after 1 and 10 years of integrated observation for each of the neutrino fluxes described above.  The energy has been divided into bins, each a factor of 10 wide.  The quantity being measured is in the total cross section averaged among events in a given bin. 

Below $\sim1$ PeV, even for the conservative Waxman-Bahcall flux, the cross section can be measured to a factor of 3 or better with only 1 year of observation.  After ten years, the accessible energy range increases to 10 PeV or higher (see figure 2).  For the optimal flux of the AMANDA-B10 limit, cross sections can be measured accurately over 100 PeV and to within one order of magnitude up to 10 EeV (see figure 3).  Figures 4 and 5 show that for a less sharply falling flux, normalized to the same number of events between 1 PeV and 1 EeV, cross sections for PeV-EeV energies are well-measurable, while cross sections at TeV energies are more challenging. Even for the most energetic colliders planned, these measurements will be impossible. For a discussion on the ability of colliders to study such effects, see Ref.\,\cite{collider}.

The systematic uncertainties involved in high energy neutrino astronomy can, presumably, be understood and limited by calibration with the atmospheric neutrino spectrum. The remaining systematic errors will result from a detector's finite angular and energy resolution.  ICECUBE is expected to achieve angular resolution below 1 degree. Also, energy resolution for shower events is expected to be at the level of 30\% or better. This is significantly more precise than the energy resolution for lepton track events.  

In conclusion, next generation neutrino telescopes may be capable of constraining the total neutrino-nucleon cross section by comparing the number of upgoing events to the number of downgoing events.  This method is independent of the shape of the neutrino flux.  Optimal energies for this measurement are in the range of 100 TeV-100 PeV where the Earth becomes opaque to neutrinos and large enough neutrino fluxes may exist for observation.  This energy range is complimentary to lower energy collider experiments and higher energy cosmic ray air shower experiments.

\begin{acknowledgments}
I would like to thank Alexander Kusenko, Jaime Alvarez-Muniz and Francis Halzen for valuable comments. This work was supported in part by  
a DOE grant No. DE-FG02-95ER40896 and in part by the Wisconsin Alumni  
Research Foundation.
\end{acknowledgments}

\vspace{1cm}

\begin{figure}[thb]
\centering\leavevmode
\includegraphics[width=3.5in]{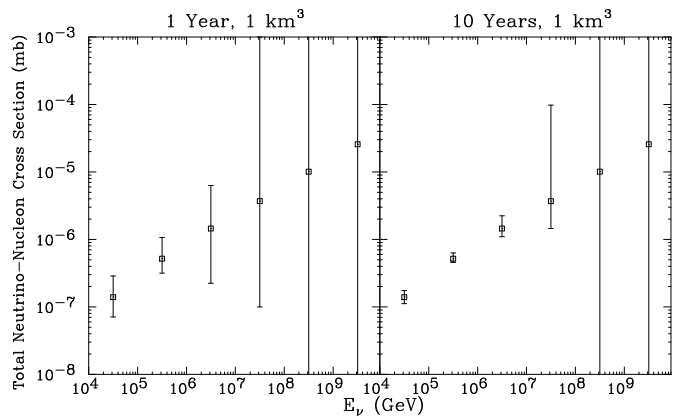}
\caption{The ability of a cubic kilometer neutrino telescope to constrain the total neutrino-nucleon cross section after 1 and 10 years observation time for the flux predicted by Waxman and Bahcall, $E_{\nu}^2 dN_{\nu}/dE_{\nu}=10^{-8}\, \rm{GeV}\, \rm{cm}^{-2}\, \rm{s}^{-1}\, \rm{sr}^{-1}$ for each of $\nu_e, \nu_{\mu}$ and $\bar{\nu_{\mu}}$ \cite{wb}.  The points are predictions based on the standard model with typical PDF extrapolations to small-x.  They have been calculated using the method described in reference \cite{neu}.}
\label{figure2}
\end{figure}

\begin{figure}[thb]
\centering\leavevmode
\includegraphics[width=3.5in]{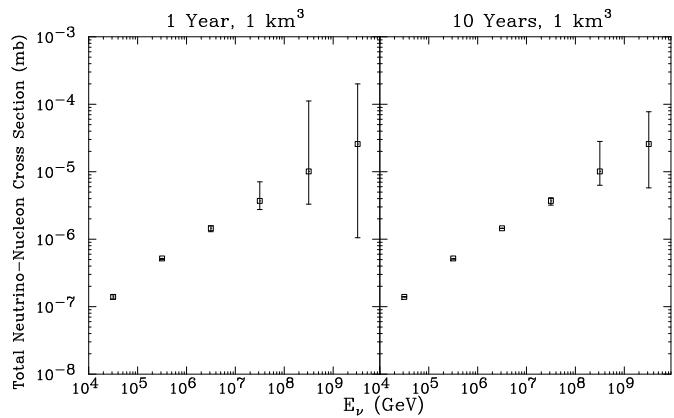}
\caption{As in figure 1, but using the flux limit from the AMANDA-B10 experiment $E_{\nu}^2 dN_{\nu}/dE_{\nu}=9 \times 10^{-7}\, \rm{GeV}\, \rm{cm}^{-2}\, \rm{s}^{-1}\, \rm{sr}^{-1}$ for each of $\nu_{\mu}$ and $\bar{\nu_{\mu}}$ \cite{amanda}.}
\label{figure3}
\end{figure}

\begin{figure}[thb]
\centering\leavevmode
\includegraphics[width=3.5in]{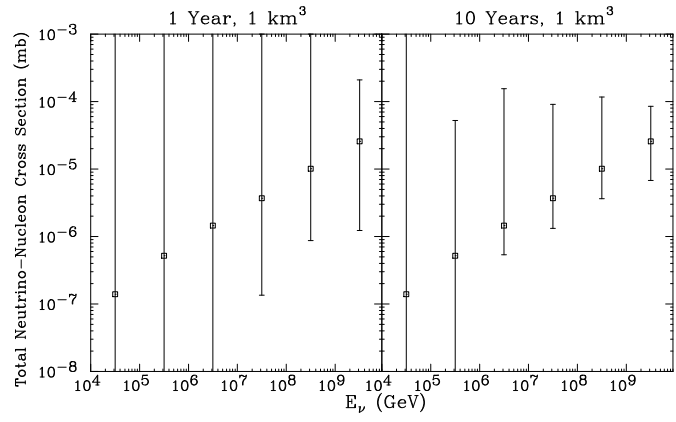}
\caption{As in figure 1, but using $E_{\nu} dN_{\nu}/dE_{\nu}=6.3 \times 10^{-12} \, \rm{cm}^{-2}\, \rm{s}^{-1}\, \rm{sr}^{-1}$ for each of $\nu_e, \nu_{\mu}$ and $\bar{\nu_{\mu}}$.  This flux predicts the same number of events between 1 PeV and 1 EeV as the Waxman-Bahcall flux.}
\label{figure4}
\end{figure}

\begin{figure}[thb]
\centering\leavevmode
\includegraphics[width=3.5in]{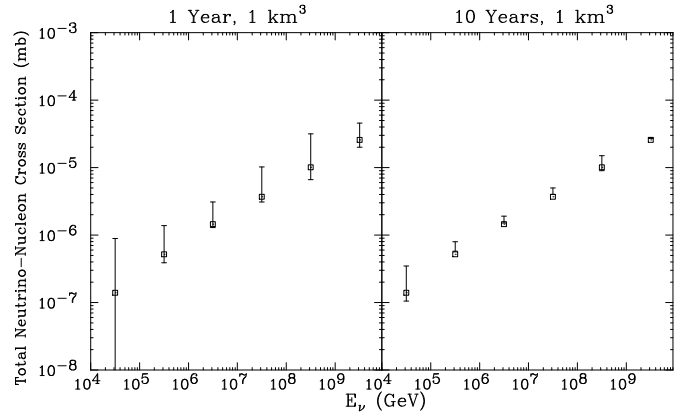}
\caption{As in figure 1, but using $E_{\nu} dN_{\nu}/dE_{\nu}=5.7 \times 10^{-10} \, \rm{cm}^{-2}\, \rm{s}^{-1}\, \rm{sr}^{-1}$ for each of $\nu_e, \nu_{\mu}$ and $\bar{\nu_{\mu}}$.  This flux predicts the same number of events between 1 PeV and 1 EeV as the AMANDA-B10 limit.}
\label{figure5}
\end{figure}

\end{document}